\documentclass[10pt]{elsart}
\usepackage{dcolumn}
\usepackage{graphicx}
\usepackage{eso-pic}
\usepackage{threeparttable}
\usepackage{cite}
\usepackage{rotating}
\usepackage{xcolor,colortbl}
\usepackage{amsmath}

\begin{document}

\begin{frontmatter}
\title{What to expect from microscopic nuclear modelling for k$_{\rm eff}$ calculations?}
\author[psi]{D. Rochman}, \author[iaea]{A. Koning}, \author[ulb]{S. Goriely} and \author[cea,saclay]{S. Hilaire}
\address[psi]{Reactor Physics and Thermal hydraulic Laboratory, Paul Scherrer Institut, Villigen, Switzerland}
\address[iaea]{Nuclear Data Section, IAEA, Wagrammerstrasse 5, 1400, Vienna, Austria}
\address[ulb]{Institut d'Astronomie et d'Astrophysique, Universit\'e Libre de Bruxelles, Campus de la Plaine, CP-226, 1050, Brussels, Belgium}
\address[cea]{CEA, DAM DIF, F-91297 Arpajon, France }
\address[saclay]{Universit\'e Paris Saclay, CEA, LMCE, F-91680 Bruy\`eres-Le-Ch\^atel, France}

\begin{abstract}
Comparisons between predicted and benchmark k$_{\rm eff}$ values from criticality-safety systems are often used as metrics to estimate 
the quality of evaluated nuclear data libraries. Relevant nuclear data  for these critical systems generally come from a mixture of expert 
knowledge and phenomenological predictions. In the present work, we use solely microscopic nuclear modelling from TALYS to estimate actinides 
cross sections and angular distributions, and we compare the calculated MCNP k$_{\rm eff}$ values for fast systems between the JEFF-3.3 
evaluated library, phenomenological and microscopic modelling. The conclusion is that even if the evaluated library leads to the most 
adequate results, the microscopic nuclear modelling can reach very similar results for these integral quantities. It demonstrates the 
remarkable advances in the recent decades of microscopic nuclear reaction ingredients for applied integral observables.
\end{abstract}

\begin{keyword}
microscopic modelling \sep TALYS \sep cross section \sep criticality
\end{keyword}

\end{frontmatter}

\section{Introduction}
Nuclear data evaluation can be performed as a combination of theoretical calculations and \textcolor{black}{local adjustements}, 
with the addition of mathematical methods based on Bayesian updates, and generalized linear least-square fitting~\cite{KONING2015207,cyrille2017,rochmanBayes}. The ingredients to these methods are generally differential data (when they exist), based on the EXFOR database~\cite{OTUKA2014272}, global or local parameter fitting (or systematics) for reaction models, and possibly integral information, in particular for nuclei of major importance in applications. Such an approach has been applied for more than 50 years, and has led to a number of general-purpose libraries~\cite{plompen2020,BROWN20181,iwamoto2023,KONING20191} as well as libraries for specific applications~\cite{fendl}. The performance of such libraries has globally improved over time, helping satisfying various users, as well as justifying this specific method of work. \\
One of the new challenges for the nuclear data community arises when using such evaluations outside their domain of validation, such as for advanced energy systems (relying on higher content of minor actinides), for higher incident energies, or for astrophysical applications. Additionally, new types of data are needed for more precise simulations ({\it e.g.} coincidence between emitted particles, more complete fission observables), and traditional nuclear reaction modelling is struggling to provide such quantities~\cite{Bernstein2019,PhysRevResearch.4.021001}. In these cases, the need for nuclear data recommendations usually goes beyond the set of stable and long-lived nuclides, which usually constitutes the bulk of evaluations included in the cited libraries. \\
A number of new developments are trying to address these limitations. One of them is to continue to improve the theoretical understanding based on the existing phenomenological prescriptions, with additional parameter adjustments, or limited improvements of specific formulas~\cite{sin2005,csige2013}. Another approach is to model specific reactions as Monte Carlo processes rather than by the use of deterministic expressions (see for instance Refs~\cite{TALOU2021108087,SCHMIDT202154}). A fundamentally different view is to model nuclear reactions based on microscopic ingredients for nuclear level densities, photon strength functions, nuclear masses, fission barriers and optical potentials~\cite{bauge2001,HILAIRE200663,GORIELY2004331,goriely2007}. Such developments requires large computational resources, but contain the potential to free oneself from {\it extrapolations of phenomenological formulas adjusted on a very narrow range of nuclei in the valley of stability}~\cite{Goriely09b,Goriely13a,goriely2018b}. They are nowadays  available for TALYS users, as the latest versions of the reaction code include ready-to-use reaction options, based on microscopic modelling~\cite{talys}. We will therefore take advantage of these new developments and apply them in the case criticality-safety benchmarks with k$_{\rm eff}$ calculations.\\
Based on the TALYS implementations of these microscopic models, the first question that this work is trying to address is the quantification of the difference for criticality calculations between predictions obtained using usual reaction models (considered as phenomenological) and advanced modelling (interpreted as microscopic). The method to answer this question is relatively simple: for a selected number of relevant nuclides, produce nuclear data (mainly cross sections and angular distributions) from these models, format them into so-called ACE files for a Monte Carlo transport code (in the present case MCNP), and use them for a selection of criticality benchmarks. The results will be analysed in a simple way, using C/E ratios, being Calculated over Experimental k$_{\rm eff}$ ratios. The distributions of these ratios can be interpreted through their  average values and standard deviations. In specific cases, values of individual criticality benchmarks will also be studied. In order to have a reference value, the calculated k$_{\rm eff}$ values based on the JEFF-3.3 library will be used. Such a comparison will potentially indicate the gain (or loss) of knowledge obtained by using such models (see section~\ref{nonadjusted}).\\
A following step in this comparison is to extend the k$_{\rm eff}$ calculations to other selected models. Many phenomenological models can be chosen (for instance among level density descriptions), and their impact can also be quantified. This part of the study will simply be performed by generalising the previous analyses to a large number of models (see section~\ref{othernonadjusted}).\\
A third and final step in this comparison is to allow for an adjustment of the model parameters in order to obtain better average C/E ratios (closer to 1) and smaller standard deviations for the distribution of the C/E values. Such a work is also very similar to the first step, as presented in section~\ref{adjusted}. This exercise will potentially indicate what can be obtained if a number of model parameters are modified, and possibly provided as new default values. The gain in terms of integral benchmarks (on k$_{\rm eff}$) will also be quantified.\\
Details are provided in the next section, and the main finding is that using default models can increase the C/E bias and standard deviation by a factor 5, but when adjusted, the C/E distribution with ingredients from microscopic model is very similar to the one of JEFF-3.3. Naturally, this work concerns only integral benchmarks, and microscopic models generally do not perform as well as a general-purpose library for differential observables such as cross sections.

\subsection{Considered reaction models}
As indicated in the introduction, the goal of the present work is to assess the impact of microscopic models for nuclear data applied on criticality calculations (see section~\ref{critical} for the description of the considered critical systems). Comparisons are performed with the JEFF-3.3 general purpose library~\cite{plompen2020}, and two types of models, namely phenomenological models on the one hand and microscopic ones on the other hand. Together with the JEFF-3.3 library, three types sets of nuclear inputs are considered.\\
The output of the models considered here cover part of the necessary aspects of the nuclear reactions, needed to perform a criticality calculation: cross sections and angular distributions in the continuum energy range. Other quantities, such as the average number of emitted neutrons per fission (called nubar), their energy spectra (called chi), and also cross sections in the resonance range, are not modified and are kept the same (and equal to JEFF-3.3) when comparing the k$_{\rm eff}$ values. The calculated cross sections of interest are therefore elastic, fission, capture and inelastic (and consequently the total cross section). For the scattering cross sections, their angular distributions are also modified and taken from the model calculations. Regarding the resonance range, although statistical resonances can be produced based on average quantities coming from different reaction models~\cite{ROCHMAN201360}, they are not modified in this work. Only cross sections in the continuum energy range (above the resolved and unresolved resonance range) are modified.\\
As mentioned in the introduction, the TALYS nuclear reaction code is used to produce such quantities, as it contains a number of models~\cite{talys}, which can simply be activated with specific keywords. In the following, two sets of models will be referenced to, simply called phenomenological and microscopic, and their characteristics are presented in Table~\ref{model}.

\begin{figure}[htbp]
\centerline{
\resizebox{0.6\columnwidth}{!}{\rotatebox{-0}{\includegraphics[trim=4cm 9cm 2cm 11cm, clip]{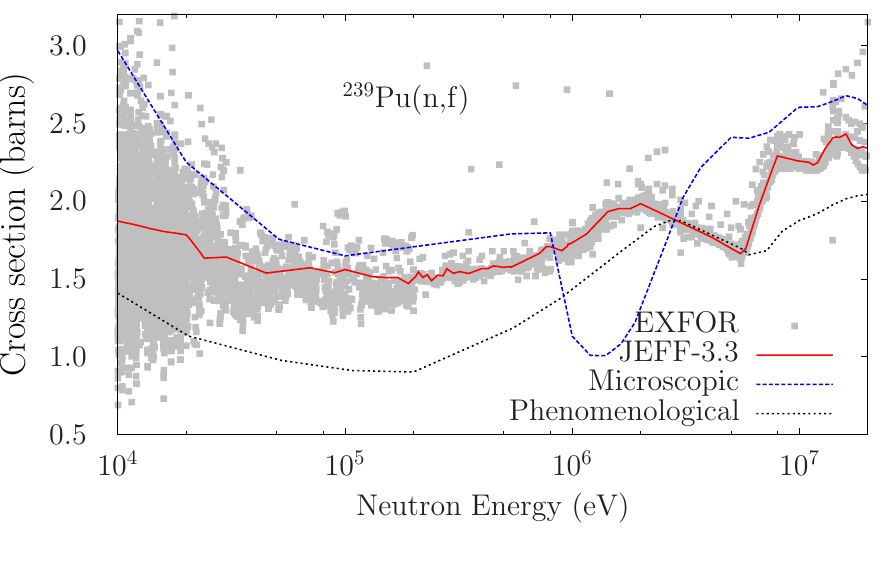}}}
}
\caption{Fission cross sections for $^{239}$Pu from the JEFF-3.3 library (adjusted to experimental data), the two different models, \textcolor{black}{and the EXFOR (measured) values}.} 
\label{fission.pu9.models} 
\end{figure}

\begin{table}[htbp]
\caption{Characteristics of the nuclear reaction models named as phenomenological and microscopic.}
\begin{center}
{\scriptsize
\begin{tabular}{l cc }
\hline
                                   & Phenomenological   &   Microscopic \\
\hline
E1 photon strength function        & Simplified Modified                 & Gogny D1M HFB+QRPA~\cite{goriely2018b} \\
                                   & Lorentzian Model (SMLO)~\cite{PLUJKO2011567}      &                               \\
Level density  (LD)                &Constant temperature & Skyrme-HFB   \\
                                   &+ Fermi gas model~\cite{PhysRevC.47.1504}    & +combinatorial~\cite{PhysRevC.78.064307} \\
Mass model                         & Experimental           & Skyrme-HFB~\cite{Goriely13a} \\
Fission barriers                   & Experimental           & Skyrme-HFB~\cite{Goriely09b}\\
Discrete levels                    & RIPL-3~\cite{Capote09} & Theoretical levels\\
Nucleus-nucleus optical potential  & local Koning-Delaroche ~\cite{KONING2003231} & JLMB~\cite{bauge2001}\\
M1 photon strength function        & \multicolumn{2}{c}{Gogny D1M HFB+QRPA~\cite{goriely2018b}}\\
Collective enhancement LD          & \multicolumn{2}{c}{Not considered}\\
Width fluctuation mode             & \multicolumn{2}{c}{Moldauer~\cite{Moldauer1980}}\\
Alpha Optical Model                & \multicolumn{2}{c}{Avrigeanu~\cite{Avrigeanu2014}}\\
\hline
\hline
\end{tabular}
}
\end{center}
\label{model}
\end{table}
In addition, the following TALYS keywords are used for the microscopic calculations: 
\begin{itemize}
    \item {\it asys y}: do not use all default level density parameters,
    \item {\it psfglobal y}: use global parametrization for the photon strength functions,
    \item {\it best n} and {\it fit n}: do not use adjusted nuclear model parameters to experimental cross sections.
\end{itemize}
For the phenomenological calculations, these options are by default opposite to these choices, {\it i.e.} experimental information is used extensively to constrain all model inputs from measured quantities.  An example of the calculated fission cross sections for $^{239}$Pu is presented in Fig.~\ref{fission.pu9.models} for the three sets of calculations.
As observed for this important reaction, differences are pronounced, but are globally contained within a factor 2. For the phenomenological models, the cross section is obtained without parameter adjustment, which is usually the starting point for improvement, based on parameter changes.  Other relevant quantities are not presented here (such as inelastic and capture cross sections), but similar range of differences are observed. Given the differences between model calculations and JEFF-3.3, the agreement with differential data as well as integral data is strongly in favor of the evaluated library. \\
Regarding the cross sections from the microscopic calculations, the agreement with the evaluated quantities has sensibly improved over the last two decades~\cite{Goriely09b}. 
In particular, mean-field calculations can now provide all the nuclear ingredients required to describe the fission path from the ground-state configuration up to the nuclear scission point. The nuclear inputs concern not only the details of the energy surface along the fission path, but also the coherent estimate of the nuclear level density derived within the combinatorial approach on the basis of the same single-particle properties, in particular at the fission saddle points. 
It was also shown in Ref.~\cite{PhysRevC.83.034601} that the various inputs can be tuned to reproduce, at best, experimental data in one unique coherent framework and be competitive with state-of-the-art data evaluations.

\subsection{Criticality benchmarks and nuclides of interest}\label{critical}
A limited number of criticality-safety benchmarks are considered in this work. In total, they correspond to 195 k$_{\rm eff}$ values 
for systems with fast spectrum, extracted from the ICSBEP database~\cite{icsbep}. 
\textcolor{black}{Other quantities such as spectral indexes can also be used, but only a small number is reported in the ICSBEP database,
compared to k$_{\rm eff}$ values. It was then decided to solely focus on criticality values.}
The MCNP6.1 neutron transport code is used for the criticality calculations~\cite{mcnp}, based on ACE files from the JEFF-3.3 library (except for the cases produced from the phenomenological and microscopic models).\\
Fast benchmarks were selected to obtain sensitivities only above the resonance range, as cross sections are not modified below the continuous region. These benchmarks are classified in four types, following the fissile contents and their concentrations, as follows:
\begin{itemize}
    \item 82 highly enriched uranium metallic systems: hmf1.1, hmf3.1, hmf3.3-hmf3.12, hmf4.1, hmf5.1-hmf5.6, hmf7.1-hmf7.9, hmf8.1, hmf9.1, hmf9.2, hmf11.1,
hmf12.1, hmf13.1, hmf15.1, hmf18.1, hmf19.1, hmf20.1, hmf21.1, hmf22.1, hmf24.1, hmf25.1-hmf25.5,
hmf26.11, hmf27.1, hmf28.1, hmf34.1-hmf34.3, hmf41.1-hmf41.6, hmf43.1-hmf43.5, hmf44.1-hmf44.5,
hmf48.1, hmf48.3, hmf48.5, hmf48.7, hmf48.9, hmf56.1, hmf57.1-hmf57.6
    \item 72 highly enriched plutonium metallic systems: pmf1.1, pmf2.1, pmf3.101-pmf3.105, pmf4.207-pmf4.215, pmf5.1, pmf6.1, pmf8.1, pmf9.1, pmf10.1, pmf11.1,
pmf12.1, pmf13.1, pmf14.1, pmf15.1, pmf16.1-pmf16.6, pmf17.201-pmf17.205, pmf18.1, pmf19.1, pmf20.1,
pmf21.1, pmf22.1, pmf23.1, pmf24.1, pmf25.1, pmf26.1, pmf27.1, pmf28.1, pmf29.1, pmf30.1, pmf31.1,
pmf32.1, pmf33.1, pmf35.1, pmf36.1, pmf37.1, pmf38.1, pmf39.1, pmf40.1, pmf41.1, pmf44.1-pmf44.5,
pmf45.1-pmf45.5, pmf46.1, pmf46.2
    \item 23 intermediate enriched uranium metallic systems: imf1.1-imf1.4, imf2.1, imf3.1, imf3.2, imf4.1, imf5.1, imf6.1, imf7.1, imf10.1, imf12.1, imf13.1, imf14.1, imf14.2, imf20.1-imf20.7, imf22.1
    \item 18 mixed uranium-plutonium metallic systems: mmf1.1, mmf2.1-mmf2.3, mmf3.1, mmf7.1, mmf7.3, mmf7.5, mmf7.9, mmf7.11, mmf7.13, mmf7.15, mmf8.1, mmf11.1-mmf11.4
\end{itemize}
These benchmarks are mostly sensitive to the $^{235,238}$U and $^{239}$Pu isotopes, but also contain other actinides such as $^{234,236}$U and $^{240,241}$Pu. In addition, there are many reflector and alloy materials made of (or containing impurities such as) Si, Cr, Cu, Fe, Ni, W, Ga, Mo, Ti, Mn and F. \\
For the calculations performed in the following, these nuclides are all considered as they are part of the benchmark definitions. For the comparison between non-adjusted reaction models (both phenomenological and microscopic, see section~\ref{nonadjusted}) and the JEFF-3.3 libraries, all the nuclear data of these 49 isotopes are changed from the JEFF-3.3 library to the selected models. For the study on the adjusted models (section~\ref{adjusted}), only the nuclear data of $^{239}$Pu is modified, and all other isotopes are kept as in JEFF-3.3.\\
Finally, given the spread of the calculated k$_{\rm eff}$ for the different models (being in the order of a few thousands of pcm), the neutron history for single MCNP simulations is set to obtain statistical uncertainties close to 50~pcm.

\section{Results}
In this section, the values of the k$_{\rm eff}$ values is presented for the different cases studied. As a large number of benchmarks are used, results are globally analysed in terms of average C/E and standard deviation for the k$_{\rm eff}$ distributions of either the 190 benchmarks (when all nuclides are considered, section~\ref{nonadjusted}), or only 90 benchmarks (when solely the $^{239}$Pu isotope is selected for the study, section~\ref{adjusted}).\\
As mentioned, once the TALYS calculations are performed for selected models (see Table~\ref{model}), the calculated cross sections and angular distributions are first formatted in an ENDF-6 file. The template for such a file is a JEFF-3.3 evaluation, where calculated cross sections and angular distributions replace the evaluated ones. This approach leads to a complete and similar ENDF-6 file (format-wise), allowing similar systematical  processing with the code NJOY into ACE formatted files. Once ACE files are obtained, the MCNP calculations are performed for the selection of benchmarks.

\subsection{Non adjusted models}\label{nonadjusted}
\begin{figure}[htbp]
\centerline{
\resizebox{0.3\columnwidth}{!}{\rotatebox{-0}{\includegraphics[trim=0cm 12cm 11cm 10cm, clip]{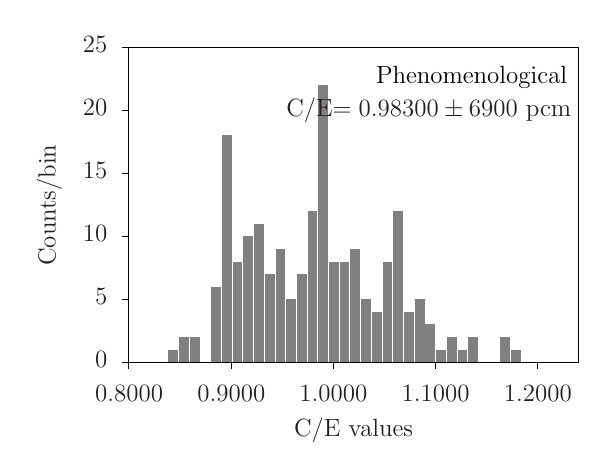}}}
\resizebox{0.3\columnwidth}{!}{\rotatebox{-0}{\includegraphics[trim=0cm 12cm 11cm 10cm, clip]{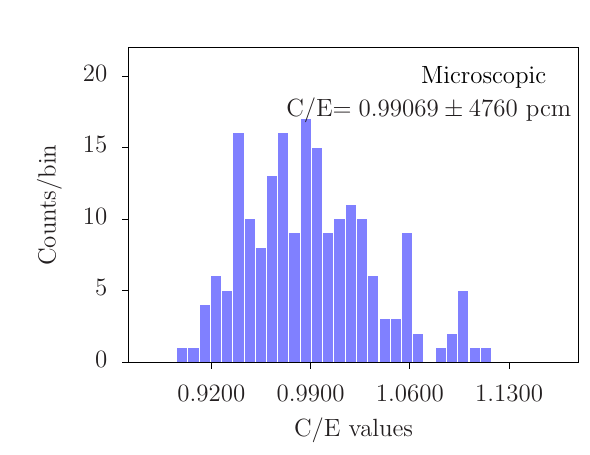}}}
\resizebox{0.3\columnwidth}{!}{\rotatebox{-0}{\includegraphics[trim=0cm 12cm 11cm 10cm, clip]{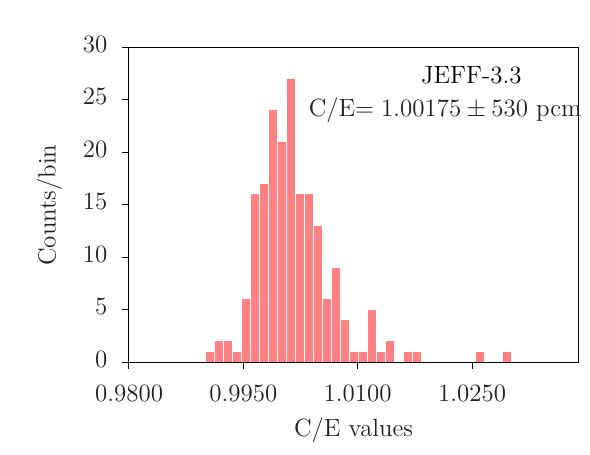}}}
}
\caption{Distribution of C/E for the 195 k$_{\rm eff}$ criticality benchmarks for the libraries and considered models.} 
\label{keff.models} 
\end{figure}

In this section, non-adjusted  models are adopted, leading to cross sections as illustrated in Fig.~\ref{fission.pu9.models}. In total, 195 benchmarks are calculated, as listed in section~\ref{critical}. Average C/E and standard deviations (1$\sigma$) are given in Table~\ref{table_stat1}, and the C/E distributions shown as histograms in Fig.~\ref{keff.models}.
\begin{table}[htbp]
\caption{Averages C/E and standard deviations for the 195 k$_{\rm eff}$ distributions coming from the non-adjusted models and the JEFF-3.3 library. Relevant cross sections and angular distributions for all actinides and structural meterials are considered.}
\begin{center}
{
\begin{tabular}{l cc }
\hline
                                   & Average C/E   &   1$\sigma$ (pcm) \\
\hline
 Phenomenological  models          &    0.98300    & 6900   \\  
 Microscopic models                &    0.99069    & 4760   \\
 JEFF-3.3                          &    1.00175    & 530 \\
\hline
\hline
\end{tabular}
}
\end{center}
\label{table_stat1}
\end{table}
As observed and expected, the JEFF-3.3 library presents the best agreement with the benchmark data, both in terms of bias and standard deviation. Such performances represent decades of developments in  nuclear data evaluation, and these dedicated efforts can be translated in the present case by a small bias (less than 200 pcm), and a relative small standard deviation (about 500 pcm). Other evaluated libraries present similar performances~\cite{iwamoto2023,BROWN20181}. Additionally, evaluated libraries are  generally performing well compared to differential data, such as cross sections compiled in the EXFOR database. \\
In the case of non-adjusted models (both phenomenological and microscopic), the biases are globally 5 to 10 times higher, with standard deviations also 10 times larger than for the evaluated libraries. 
It can be noted that usually phenomenological models can lead to better results, but in the present case, their parameters were not fine tuned but rather obtained globally from systematics. In the microscopic case, the model ingredients are even not renormalized to or replaced by experimental values, in particular purely theoretical HFB masses are considered with no resort to known masses. Such differences in averages and standard deviations can be interpreted as the effect of dedicated efforts within the evaluation community to obtain good criticality performances, compared to model developments which are not focusing on these benchmarks and are considering different information for theoretical improvements. \\
An unexpected observation is that microscopic models lead to slightly better average C/E compared to phenomenological models. This can be interpreted as one of the advantages to use predictions  based on microscopic nuclear inputs, especially when no evaluated data is available. 

\subsection{Other non-adjusted models}\label{othernonadjusted}
\begin{figure}[htbp]
\centerline{
\resizebox{0.4\columnwidth}{!}{\rotatebox{-0}{\includegraphics[trim=0cm 12cm 10cm 10cm, clip]{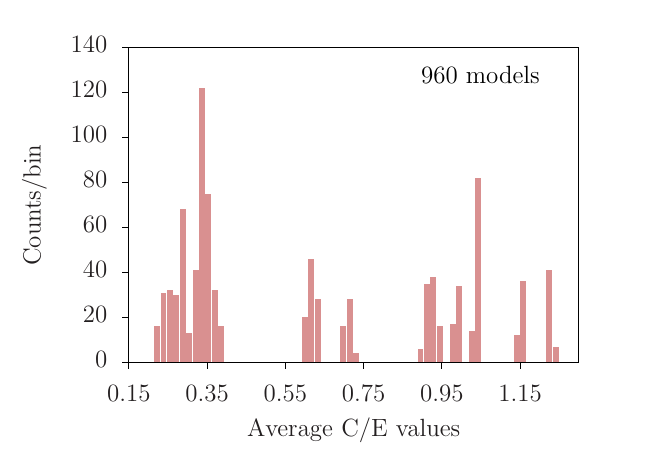}}}
\resizebox{0.4\columnwidth}{!}{\rotatebox{-0}{\includegraphics[trim=0cm 12cm 10cm 10cm, clip]{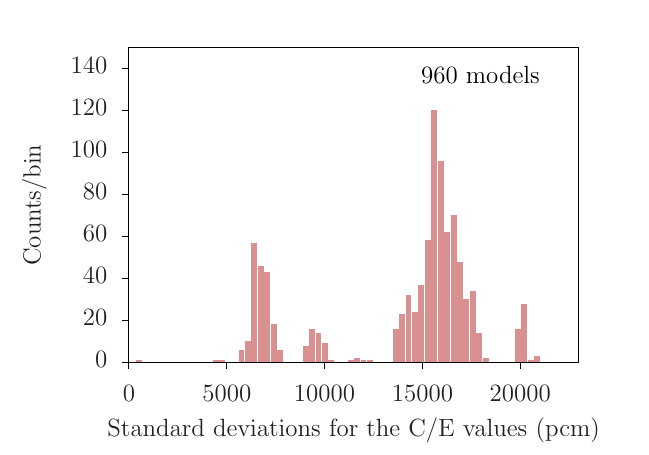}}}
}
\caption{Distributions of the average C/E values (left) and standard deviations (right) for the 960 model combinations, with 195 k$_{\rm eff}$ criticality benchmarks for each combination.} 
\label{keff.allmodels} 
\end{figure}

Following the selection of models given in Table~\ref{model}, one can generalize the model variations, for instance by selecting different mass models, level densities, or photon strength functions. This type of approach was already performed in TENDL-2023, for astrophysical rates~\cite{rochman2024}, where combinations of various inputs were considered. In total, for non-fissile nuclides, combinations of level densities models, photon strength function models and others has led to 480 different possibilities; for fissile nuclides, as two fission models were considered (experimental fission barriers and WKB approximations for fission path model~\cite{PhysRevC.74.014608}), the number of combinations rose to 960. \\
As this represents a large number of TALYS calculations (for 49 isotopes from section~\ref{critical}), only 7 isotopes were selected, covering the majority of the effect (being all U and Pu actinides). Whereas the variation of the 49 isotopes for the microscopic model gave an average C/E of 0.99069 ($\pm$ 4760 pcm), the 7 actinides correspond to 0.99233 ($\pm$ 4560 pcm), being close enough to the total effect. Based on these 7 actinides, all the 195 benchmarks were calculated again, this time for each of the 960 model combinations. Details of the models can be found in Ref.~\cite{rochman2024}, corresponding to a mixture of microscopic and phenomenological models, excluding a number of nonphysical or incoherent possibilities. The histograms of the distributions for the average C/E and standard deviation values are presented in Fig.~\ref{keff.allmodels}.
In the left histogram, one count in a bin is the average of the 195 C/E values for one of the 960 model combinations. In total, there are 960 counts (or models) in the histogram. Similarly, the right histogram presents the 960 standard deviation values.
Some of these model combinations systematically provide very low or very high k$_{\rm eff}$ values 
(and consequently average C/E values).  
\textcolor{black}{In the future, these combinations should not be considered as they are leading to extreme C/E values.}
In total, the spreads of the average and standard deviation values are very broad, which provides a perspective on the performances of the default phenomenological and microscopic models, as presented in Table~\ref{table_stat1}.  If the model values from Table~\ref{table_stat1} can be interpreted as being of poor performances with respect to the JEFF-3.3 library, they are in fact performing well compared to other model combinations. \\
In the next and last section, we focus on the possible improvement for the microscopic models, starting from the choices presented in Table~\ref{model}.

\subsection{Adjusted models for $^{239}$Pu based on criticality benchmarks}\label{adjusted}
\begin{figure}[htbp]
\centerline{
\resizebox{0.3\columnwidth}{!}{\rotatebox{-0}{\includegraphics[trim=0cm 12cm 11cm 10cm, clip]{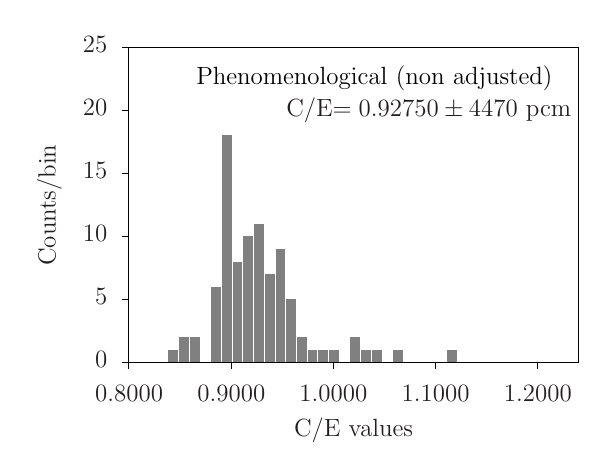}}}
\resizebox{0.3\columnwidth}{!}{\rotatebox{-0}{\includegraphics[trim=0cm 12cm 11cm 10cm, clip]{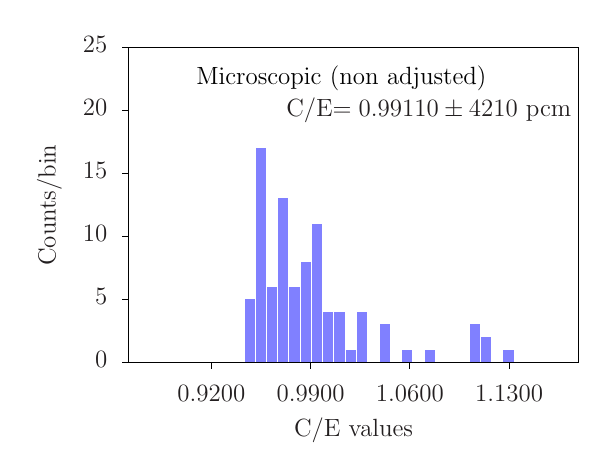}}}
}
\centerline{
\resizebox{0.3\columnwidth}{!}{\rotatebox{-0}{\includegraphics[trim=0cm 12cm 11cm 10cm, clip]{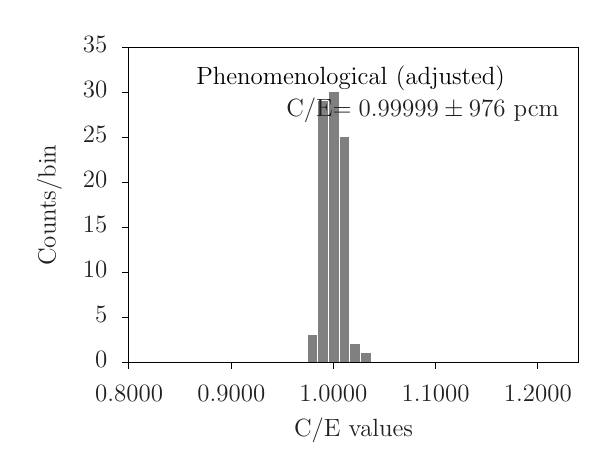}}}
\resizebox{0.3\columnwidth}{!}{\rotatebox{-0}{\includegraphics[trim=0cm 12cm 11cm 10cm, clip]{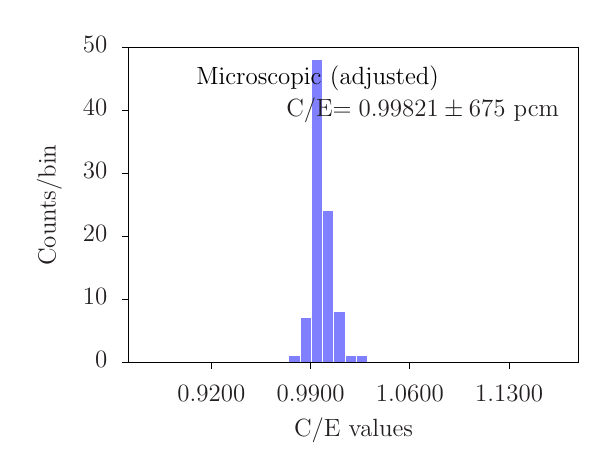}}}
}
\caption{Distribution of C/E for the 90 $^{239}$Pu k$_{\rm eff}$ criticality benchmarks for the libraries and considered models. Top: non adjusted calculations; bottom: with adjusted parameters for the microscopic and phenomenological models. } 
\label{keff.pu9.models} 
\end{figure}

\begin{figure}[htbp]
\centerline{
\resizebox{0.6\columnwidth}{!}{\rotatebox{-0}{\includegraphics[trim=4cm 9cm 2cm 11cm, clip]{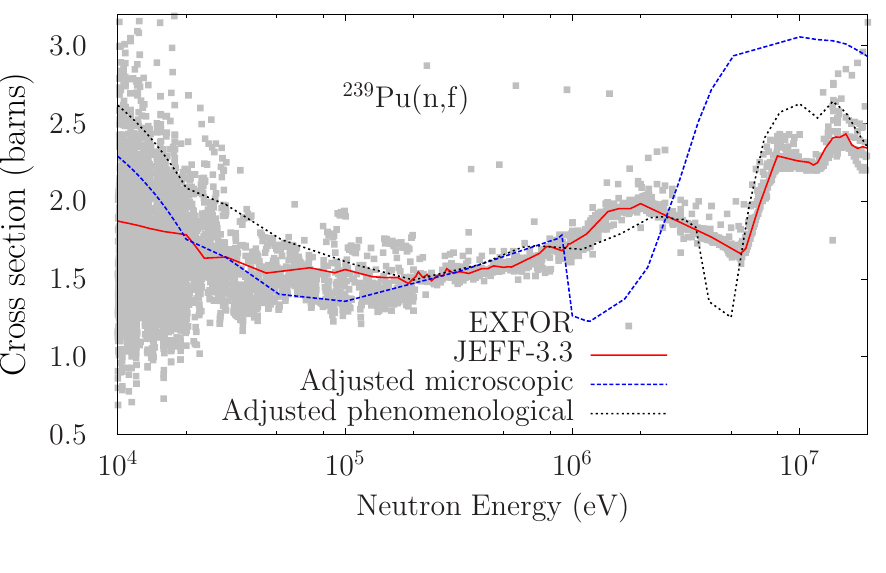}}}
}
\caption{Fission cross sections for $^{239}$Pu from the JEFF-3.3 library, \textcolor{black}{EXFOR} and the two different  models adjusted on criticality benchmarks.} 
\label{fission.pu9.adjusted.models} 
\end{figure}

\begin{figure}[htbp]
\centerline{
\resizebox{0.95\columnwidth}{!}{\rotatebox{-0}{\includegraphics[trim=0cm 13cm 0cm 8cm, clip]{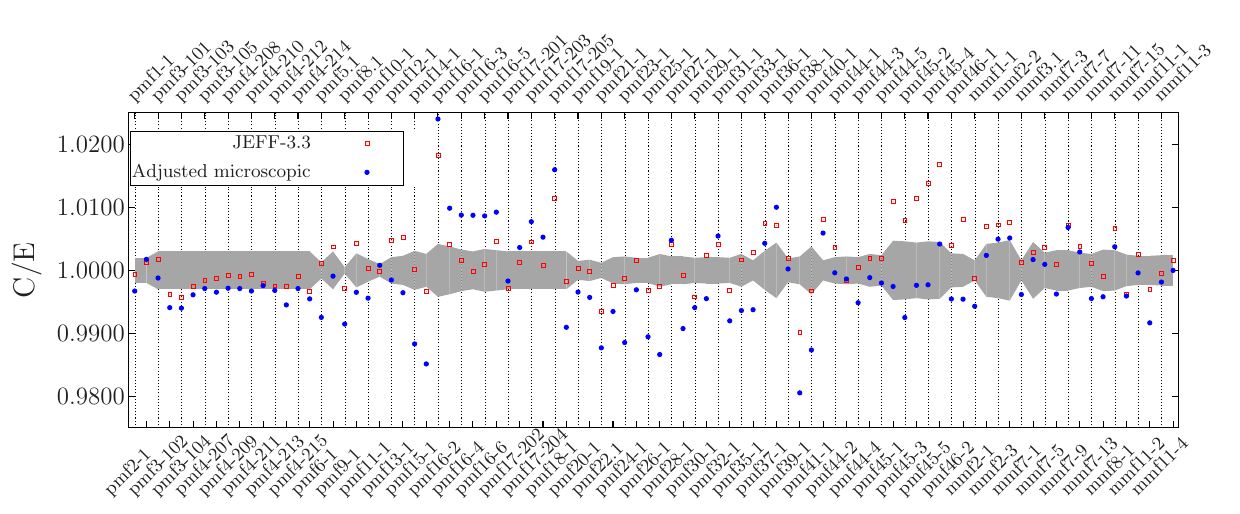}}}
}
\caption{Values of the C/E for each of the 90 benchmarks for which $^{239}$Pu is either from JEFF-3.3 or from the adjusted microscopic models (other isotopes are from JEFF-3.3). } 
\label{benchmarks1} 
\end{figure}

The global results presented in Table~\ref{table_stat1} can be interpreted in different ways, depending on the reference value. With respect to a library such as JEFF-3.3, performances indicate biases 5 to 10 times higher (as well as for the spread), even if, surprisingly, the microscopic models lead to better agreement than the phenomenological models. On the other hand, these two sets of models perform well compared to other selections, as shown in Fig.~\ref{keff.allmodels}.\\
If general-purpose libraries are already in better agreement with differential data than the considered models, it has been argued that such libraries were to some extent adjusted to integral benchmarks (for instance with the criticality benchmarks used in this work). This was performed not based on model parameters, but rather on local modifications (at specific energies) of cross sections, nubar, energy and possibly angular distributions. In the present approach based on models, possible adjustments can only be done with model parameters, but it can certainly be performed as efficiently as in the case of a library, given that only a limited number of observables are considered. \\
In the present work, such model parameter adjustment was performed for $^{239}$Pu only. From the 195 selected benchmarks, 90 are  sensitive to $^{239}$Pu, mainly the pmf and mmf benchmarks, as listed in section~\ref{critical}. The procedure of adjustment is relatively simple and was described in Refs.~\cite{rochman2011,rochman2022}: sample model parameters, produce corresponding sampled cross sections, and use them in criticality calculations. The parameters leading to average k$_{\rm eff}$ C/E values close to 1 are then selected. In the case of the phenomenological models, the sampling of model parameters is performed as in the case of the TENDL library~\cite{KONING20191}, based on the TALYS and TASMAN codes. For microscopic models, the same approach is followed, but a selection of 17 relevant parameters, as described in the TALYS manual, is performed. Results of the selection of the model parameters in terms of average C/E for the 90 k$_{\rm eff}$ benchmarks are presented in Table~\ref{table_stat2} and Fig.~\ref{keff.pu9.models}.
\begin{table}[htbp]
\caption{Averages C/E and standard deviations for the distributions of the 90 $^{239}$Pu k$_{\rm eff}$  coming from the non-adjusted and adjusted models, as well as from the JEFF-3.3 library.}
\begin{center}
{
\begin{tabular}{l cc }
\hline
                                   & Average C/E   &   1$\sigma$ (pcm) \\
\hline
 Non adjusted phenomenological (default) models  &    0.92750    & 4470   \\  
 Non adjusted microscopic models                 &    0.99110    & 4210   \\
 Adjusted phenomenological  models               &    0.99999    &  976   \\  
 Adjusted Microscopic models                     &    0.99821    &  675   \\
 JEFF-3.3                                        &    1.00179    &  480   \\
\hline
\hline
\end{tabular}
}
\end{center}
\label{table_stat2}
\end{table}
The modification of the model parameters allows to significantly reduce the average biases for both model types, as well as to strongly reduce the spread (standard deviation) of the C/E values. The adjusted results are of the same order of magnitude than for JEFF-3.3. 
As mentioned, this adjustment only considers the integral information of the criticality benchmarks, and is not bounded by differential data for the relevant cross sections and angular distributions. As seen in Fig.~\ref{fission.pu9.models}, the non adjusted cross sections (or prior) are within a factor 2 with respect to the JEFF-3.3 library. The adjusted fission cross sections, as presented in Fig.~\ref{fission.pu9.adjusted.models}, are also relatively close to the JEFF-3.3 values (within a factor 2 as well), allowing better k$_{\rm eff}$ C/E values.
It is very likely that compensations between high and low fission cross sections (as a function of energy) help to improve the global agreement with the integral benchmarks. As many degrees of freedom are available with the phenomenological modelling, the shape of plotted cross section can certainly be improved; on the contrary, the possibilities are more limited in the case of the microscopic calculations, and it is for instance very difficult and beyond the scope of the current study to increase the fission
cross section at the minima (close to 1.5 MeV) and at the same time decrease it above 4 MeV. \\
Nonetheless, the bias, and more remarkably the standard deviation in the case of the adjusted microscopic calculations can be strongly reduced: to less than 200~pcm for the bias (given a statistical uncertainty for each k$_{\rm eff}$ value of about 50~pcm), and by a factor 5 for the standard deviation (rendering it similar to the one of the JEFF-3.3 library). \\
As an additional remark on the use of microscopic and phenomenological models, a combination of both approaches can lead to good C/E values, as well as improved agreement with the JEFF-3.3 fission cross section: if one uses  microscopic models except for masses (experimental masses instead of the Skyrme-HFB) and for the optical potential Sukhovitskii model~\cite{sukho2000} instead of the JLMB model, the average C/E and sigma are 1.00002 $\pm$ 719~pcm, while reducing the difference with the JEFF-3.3 fission cross section well within a factor 1.5. This demonstrates the importance of specific models for criticality benchmarks and
the possibility to significantly improve differential cross sections.\\
Individual C/E values for the 90 considered benchmarks are presented in Fig.~\ref{benchmarks1}, for both the JEFF-3.3 library and with the adjusted $^{239}$Pu nuclear data.
Even if averages and standard deviations are similar, differences can be observed. Additionally, the pmf benchmarks are strongly correlated due to the variations of the $^{239}$Pu, based on the microscopic models. The correlations are generally higher than +0.99, which implies that all the calculated k$_{\rm eff}$ for these benchmarks can only increase or decrease together, due to the variations of $^{239}$Pu cross sections (and angular distributions). The rigidity of the $^{239}$Pu cross section shape, when varying the parameters of the microscopic models, is too strong to modify the calculated k$_{\rm eff}$ values by different quantities. Similarly, the correlation terms between pmf and mmf benchmarks are very high as well ($>+0.94$), implying behaviour alike as between the pmf benchmarks themselves. \\
Consequently, the current state of the microscopic modelling already leads to results of significant quality with respect to integral benchmarks (Fig.~\ref{benchmarks1}), but an excess of  cross section hardness does not allow local adjustments. Of possibly higher importance, the cross section shapes are still within a factor 2 when comparing with evaluated or measured differential values (Fig.~\ref{fission.pu9.adjusted.models}), indicating possible range of improvement.

\section{Conclusion}
The main goal of this work was to quantify the impact of the recent microscopic nuclear inputs for cross sections and angular distributions 
on integral criticality benchmarks with fast  neutron spectra. Relevant nuclear data for the benchmark cores and possible reflectors were 
calculated with TALYS and the k$_{\rm eff}$  values were obtained with MCNP6. It was shown that the bias is about 5 times larger for the 
average C/E than for the evaluated JEFF-3.3 library, and 10 times larger for the spread (standard deviation), see Table~\ref{table_stat1}. 
Compared to other existing reaction models, such performances are among the best, and surprisingly good when keeping in mind that for the 
microscopic nuclear inputs, no experimental information is directly introduced in the calculation. After adjustment of the model parameters, 
the bias and standard deviation obtained from the microscopic models become similar to those of the JEFF-3.3 library (see Table~\ref{table_stat2}). 
This study indicates the tremendous progress of the microscopic nuclear modelling in terms of integral validation, while it was also observed that
 the shape of the obtained cross sections significantly differs from the evaluated ones. It should be stressed that these microscopic models were 
based on theoretical masses, discrete levels and nuclear level densities, not normalized to observed quantities ({\it e.g.} average level spacings, 
low-lying states). Future model developments will certainly keep on improving the predictions of such observables.

\section*{Acknowledgements}
This work was partly funded by the European Union's Horizon 2020 Research and Innovation
Programme under grant agreement No 847552  (project SANDA, Supplying Accurate Nuclear Data for energy and non-energy Applications) 
and 101164596 (project APRENDE, Addressing PRiorities of Evaluated Nuclear Data in Europe). 
This work was supported by the Fonds de la Recherche Scientifique (F.R.S.-FNRS) and the Fonds Wetenschappelijk Onderzoek - Vlaanderen (FWO) under the EOS Projects nr O022818F and O000422F and by the European Union (ChECTEC-INFRA, project no. 101008324). 

\newpage
\bibliographystyle{ieeetr}
\bibliography{bibliography.bib}

\end{document}